\documentclass{svmult}
\usepackage{epsfig,graphicx} 
\usepackage[bottom]{footmisc}
\usepackage{natbib,url}      
\bibpunct{(}{)}{;}{a}{}{,}   
\def\cite#1{\citealp{#1}}    
\def\authorindex#1{}  
\def\figspath{.}  

\begin{document}\newcount\preprintheader\preprintheader=1



\def\thisvolume{these proceedings}

\def\aj{{AJ}}			
\def\araa{{ARA\&A}}		
\def\apj{{ApJ}}			
\def\apjl{{ApJ}}		
\def\apjs{{ApJS}}		
\def\ao{{Appl.\ Optics}} 
\def\apss{{Ap\&SS}}		
\def\aap{{A\&A}}		
\def\aapr{{A\&A~Rev.}}		
\def\aaps{{A\&AS}}		
\def\an{{Astron.\ Nachrichten}}
\def\aspcs{{ASP Conf.\ Ser.}}
\def\assp{{Astrophys.\ \& Space Sci.\ Procs., Springer, Heidelberg}}
\def\azh{{AZh}}			
\def\baas{{BAAS}}		
\def\jrasc{{JRASC}}	
\def\memras{{MmRAS}}		
\def\mnras{{MNRAS}}
\def\nat{{Nat}}		
\def\pra{{Phys.\ Rev.\ A}} 
\def\prb{{Phys.\ Rev.\ B}}		
\def\prc{{Phys.\ Rev.\ C}}		
\def\prd{{Phys.\ Rev.\ D}}		
\def\prl{{Phys.\ Rev.\ Lett.}} 
\def\pasp{{PASP}}
\def\pasj{{PASJ}}		
\def\qjras{{QJRAS}}
\def\science{{Sci}}		
\def\skytel{{S\&T}}		
\def\solphys{{Solar\ Phys.}} 
\def\sovast{{Soviet\ Ast.}}  
\def\ssr{{Space\ Sci.\ Rev.}}
\def\svassp{{Astrophys.\ Space Sci.\ Procs., Springer, Heidelberg}}
\def\zap{{ZAp}}			
\let\astap=\aap
\let\apjlett=\apjl
\let\apjsupp=\apjs
\def\grl{{Geophys.\ Res.\ Lett.}}  
\def\jgr{{J. Geophys.\ Res.}} 

\def\ion#1#2{{\rm #1}\,{\uppercase{#2}}}  
\def\deg{\hbox{$^\circ$}}
\def\sun{\hbox{$\odot$}}
\def\earth{\hbox{$\oplus$}}
\def\la{\mathrel{\hbox{\rlap{\hbox{\lower4pt\hbox{$\sim$}}}\hbox{$<$}}}}
\def\ga{\mathrel{\hbox{\rlap{\hbox{\lower4pt\hbox{$\sim$}}}\hbox{$>$}}}}
\def\sq{\hbox{\rlap{$\sqcap$}$\sqcup$}}
\def\arcmin{\hbox{$^\prime$}}
\def\arcsec{\hbox{$^{\prime\prime}$}}
\def\fd{\hbox{$.\!\!^{\rm d}$}}
\def\fh{\hbox{$.\!\!^{\rm h}$}}
\def\fm{\hbox{$.\!\!^{\rm m}$}}
\def\fs{\hbox{$.\!\!^{\rm s}$}}
\def\fdg{\hbox{$.\!\!^\circ$}}
\def\farcm{\hbox{$.\mkern-4mu^\prime$}}
\def\farcs{\hbox{$.\!\!^{\prime\prime}$}}
\def\fp{\hbox{$.\!\!^{\scriptscriptstyle\rm p}$}}
\def\micron{\hbox{$\mu$m}}
\def\onehalf{\hbox{$\,^1\!/_2$}}	
\def\onethird{\hbox{$\,^1\!/_3$}}
\def\twothirds{\hbox{$\,^2\!/_3$}}
\def\onequarter{\hbox{$\,^1\!/_4$}}
\def\threequarters{\hbox{$\,^3\!/_4$}}
\def\ubv{\hbox{$U\!BV$}}		
\def\ubvr{\hbox{$U\!BV\!R$}}		
\def\ubvri{\hbox{$U\!BV\!RI$}}		
\def\ubvrij{\hbox{$U\!BV\!RI\!J$}}		
\def\ubvrijh{\hbox{$U\!BV\!RI\!J\!H$}}		
\def\ubvrijhk{\hbox{$U\!BV\!RI\!J\!H\!K$}}		
\def\ub{\hbox{$U\!-\!B$}}		
\def\bv{\hbox{$B\!-\!V$}}		
\def\vr{\hbox{$V\!-\!R$}}		
\def\ur{\hbox{$U\!-\!R$}}


\def\labelitemi{{\bf --}}  

\def\rmit#1{{\it #1}}              
\def\rmit#1{{\rm #1}}              
\def\etal{\rmit{et al.}}           
\def\etc{\rmit{etc.}}           
\def\ie{\rmit{i.e.,}}              
\def\eg{\rmit{e.g.,}}              
\def\cf{cf.}                       
\def\viz{\rmit{viz.}}
\def\vs{\rmit{vs.}}

\def\rot{\hbox{\rm rot}}
\def\div{\hbox{\rm div}}
\def\lesssim{\mathrel{\hbox{\rlap{\hbox{\lower4pt\hbox{$\sim$}}}\hbox{$<$}}}}
\def\gtrsim{\mathrel{\hbox{\rlap{\hbox{\lower4pt\hbox{$\sim$}}}\hbox{$>$}}}}
\def\mathstacksym#1#2#3#4#5{\def#1{\mathrel{\hbox to 0pt{\lower 
    #5\hbox{#3}\hss} \raise #4\hbox{#2}}}}
\mathstacksym\lesssim{$<$}{$\sim$}{1.5pt}{3.5pt} 
\mathstacksym\gtrsim{$>$}{$\sim$}{1.5pt}{3.5pt} 
\mathstacksym\lrarrow{$\leftarrow$}{$\rightarrow$}{2pt}{1pt} 
\mathstacksym\lessgreat{$>$}{$<$}{3pt}{3pt} 

\def\dif{\: {\rm d}}                       
\def\ep{\:{\rm e}^}                        
\def\dash{\hbox{$\,-\,$}}                  
\def\is{\!=\!}                             

\def\starname#1#2{${#1}$\,{\rm {#2}}}  
\def\Teff{\hbox{$T_{\rm eff}$}}   

\def\kms{\hbox{km$\;$s$^{-1}$}}
\def\ms{\hbox{m$\;$s$^{-1}$}}
\def\Mxcm{\hbox{Mx\,cm$^{-2}$}}    

\def\Bapp{\hbox{$B_{\rm app}$}}    

\def\komega{($k, \omega$)}                 
\def\kf{($k_h,f$)}                         
\def\VminI{\hbox{$V\!\!-\!\!I$}}           
\def\IminI{\hbox{$I\!\!-\!\!I$}}           
\def\VminV{\hbox{$V\!\!-\!\!V$}}           
\def\Xt{\hbox{$X\!\!-\!t$}}                

\def\level #1 #2#3#4{$#1 \: ^{#2} \mbox{#3} ^{#4}$}   

\def\specchar#1{\uppercase{#1}}    
\def\AlI{\mbox{Al\,\specchar{i}}}  
\def\BI{\mbox{B\,\specchar{i}}} 
\def\BII{\mbox{B\,\specchar{ii}}}  
\def\BaI{\mbox{Ba\,\specchar{i}}}  
\def\BaII{\mbox{Ba\,\specchar{ii}}} 
\def\CI{\mbox{C\,\specchar{i}}} 
\def\CII{\mbox{C\,\specchar{ii}}} 
\def\CIII{\mbox{C\,\specchar{iii}}} 
\def\CIV{\mbox{C\,\specchar{iv}}} 
\def\CaI{\mbox{Ca\,\specchar{i}}} 
\def\CaII{\mbox{Ca\,\specchar{ii}}} 
\def\CaIII{\mbox{Ca\,\specchar{iii}}} 
\def\CoI{\mbox{Co\,\specchar{i}}} 
\def\CrI{\mbox{Cr\,\specchar{i}}} 
\def\CriI{\mbox{Cr\,\specchar{ii}}} 
\def\CsI{\mbox{Cs\,\specchar{i}}} 
\def\CsII{\mbox{Cs\,\specchar{ii}}} 
\def\CuI{\mbox{Cu\,\specchar{i}}} 
\def\FeI{\mbox{Fe\,\specchar{i}}} 
\def\FeII{\mbox{Fe\,\specchar{ii}}} 
\def\FeIX{\mbox{Fe\,\specchar{ix}}}
\def\FeX{\mbox{Fe\,\specchar{x}}}
\def\FeXVI{\mbox{Fe\,\specchar{xvi}}}
\def\FrI{\mbox{Fr\,\specchar{i}}}
\def\HI{\mbox{H\,\specchar{i}}} 
\def\HII{\mbox{H\,\specchar{ii}}} 
\def\Hmin{\hbox{\rmH$^{^{_{\scriptstyle -}}}$}}      
\def\Hemin{\hbox{{\rm He}$^{^{_{\scriptstyle -}}}$}} 
\def\HeI{\mbox{He\,\specchar{i}}} 
\def\HeII{\mbox{He\,\specchar{ii}}} 
\def\HeIII{\mbox{He\,\specchar{iii}}} 
\def\KI{\mbox{K\,\specchar{i}}} 
\def\KII{\mbox{K\,\specchar{ii}}} 
\def\KIII{\mbox{K\,\specchar{iii}}} 
\def\LiI{\mbox{Li\,\specchar{i}}} 
\def\LiII{\mbox{Li\,\specchar{ii}}} 
\def\LiIII{\mbox{Li\,\specchar{iii}}} 
\def\MgI{\mbox{Mg\,\specchar{i}}} 
\def\MgII{\mbox{Mg\,\specchar{ii}}} 
\def\MgIII{\mbox{Mg\,\specchar{iii}}} 
\def\MnI{\mbox{Mn\,\specchar{i}}} 
\def\NI{\mbox{N\,\specchar{i}}}
\def\NIV{\mbox{N\,\specchar{iv}}}
\def\NaI{\mbox{Na\,\specchar{i}}}
\def\NaII{\mbox{Na\,\specchar{ii}}}
\def\NaIII{\mbox{Na\,\specchar{iii}}}
\def\NeVIII{\mbox{Ne\,\specchar{viii}}} 
\def\NiI{\mbox{Ni\,\specchar{i}}} 
\def\NiII{\mbox{Ni\,\specchar{ii}}}
\def\NiIII{\mbox{Ni\,\specchar{iii}}} 
\def\OI{\mbox{O\,\specchar{i}}} 
\def\OVI{\mbox{O\,\specchar{vi}}}
\def\RbI{\mbox{Rb\,\specchar{i}}} 
\def\SII{\mbox{S\,\specchar{ii}}} 
\def\SiI{\mbox{Si\,\specchar{i}}} 
\def\SiII{\mbox{Si\,\specchar{ii}}} 
\def\SrI{\mbox{Sr\,\specchar{i}}}
\def\SrII{\mbox{Sr\,\specchar{ii}}}
\def\TiI{\mbox{Ti\,\specchar{i}}} 
\def\TiII{\mbox{Ti\,\specchar{ii}}} 
\def\TiIII{\mbox{Ti\,\specchar{iii}}} 
\def\TiIV{\mbox{Ti\,\specchar{iv}}} 
\def\VI{\mbox{V\,\specchar{i}}} 
\def\HtwoO{\mbox{H$_2$O}}        
\def\Otwo{\mbox{O$_2$}}          

\def\Halpha{\mbox{H\hspace{0.1ex}$\alpha$}} 
\def\Ha{\mbox{H\hspace{0.2ex}$\alpha$}}
\def\Hbeta{\mbox{H\hspace{0.2ex}$\beta$}}
\def\Hgamma{\mbox{H\hspace{0.2ex}$\gamma$}}
\def\Hdelta{\mbox{H\hspace{0.2ex}$\delta$}}
\def\Hepsilon{\mbox{H\hspace{0.2ex}$\epsilon$}}
\def\Hzeta{\mbox{H\hspace{0.2ex}$\zeta$}}
\def\Lyalpha{\mbox{Ly$\hspace{0.2ex}\alpha$}}
\def\Lybeta{\mbox{Ly$\hspace{0.2ex}\beta$}}
\def\Lygamma{\mbox{Ly$\hspace{0.2ex}\gamma$}}
\def\Lycont{\mbox{Ly\hspace{0.2ex}{\small cont}}}
\def\Baalpha{\mbox{Ba$\hspace{0.2ex}\alpha$}}
\def\Babeta{\mbox{Ba$\hspace{0.2ex}\beta$}}
\def\Bacont{\mbox{Ba\hspace{0.2ex}{\small cont}}}
\def\Paalpha{\mbox{Pa$\hspace{0.2ex}\alpha$}}
\def\Bralpha{\mbox{Br$\hspace{0.2ex}\alpha$}}

\def\NaD{\mbox{Na\,\specchar{i}\,D}}    
\def\NaDone{\mbox{Na\,\specchar{i}\,\,D$_1$}}
\def\NaDtwo{\mbox{Na\,\specchar{i}\,\,D$_2$}}
\def\NaID{\mbox{Na\,\specchar{i}\,\,D}}
\def\NaIDone{\mbox{Na\,\specchar{i}\,\,D$_1$}}
\def\NaIDtwo{\mbox{Na\,\specchar{i}\,\,D$_2$}}
\def\Done{\mbox{D$_1$}}
\def\Dtwo{\mbox{D$_2$}}

\def\Mgbone{\mbox{Mg\,\specchar{i}\,b$_1$}}
\def\Mgbtwo{\mbox{Mg\,\specchar{i}\,b$_2$}}
\def\Mgbthree{\mbox{Mg\,\specchar{i}\,b$_3$}}
\def\MgIb{\mbox{Mg\,\specchar{i}\,b}}
\def\MgIbone{\mbox{Mg\,\specchar{i}\,b$_1$}}
\def\MgIbtwo{\mbox{Mg\,\specchar{i}\,b$_2$}}
\def\MgIbthree{\mbox{Mg\,\specchar{i}\,b$_3$}}

\def\CaIIK{\mbox{Ca\,\specchar{ii}\,K}}       
\def\CaIIH{\mbox{Ca\,\specchar{ii}\,H}}
\def\CaIIHK{\mbox{Ca\,\specchar{ii}\,H\,\&\,K}}
\def\HK{\mbox{H\,\&\,K}}
\def\Kthree{\mbox{K$_3$}}      
\def\Hthree{\mbox{H$_3$}}
\def\Ktwo{\mbox{K$_2$}}
\def\Htwo{\mbox{H$_2$}}
\def\Kone{\mbox{K$_1$}}     
\def\Hone{\mbox{H$_1$}}     
\def\KtwoV{\mbox{K$_{2V}$}}
\def\KtwoR{\mbox{K$_{2R}$}}
\def\KoneV{\mbox{K$_{1V}$}}
\def\KoneR{\mbox{K$_{1R}$}}
\def\HtwoV{\mbox{H$_{2V}$}}
\def\HtwoR{\mbox{H$_{2R}$}}
\def\HoneV{\mbox{H$_{1V}$}}
\def\HoneR{\mbox{H$_{1R}$}}

\def\hk{\mbox{h\,\&\,k}}
\def\kthree{\mbox{k$_3$}}    
\def\hthree{\mbox{h$_3$}}
\def\ktwo{\mbox{k$_2$}}
\def\htwo{\mbox{h$_2$}}
\def\kone{\mbox{k$_1$}}     
\def\hone{\mbox{h$_1$}}     
\def\ktwoV{\mbox{k$_{2V}$}}
\def\ktwoR{\mbox{k$_{2R}$}}
\def\koneV{\mbox{k$_{1V}$}}
\def\koneR{\mbox{k$_{1R}$}}
\def\htwoV{\mbox{h$_{2V}$}}
\def\htwoR{\mbox{h$_{2R}$}}
\def\honeV{\mbox{h$_{1V}$}}
\def\honeR{\mbox{h$_{1R}$}}

\ifnum\preprintheader=1     
\makeatletter  
\def\@maketitle{\newpage
\markboth{}{}%
  {\mbox{} \vspace*{-8ex} \par 
   \em \footnotesize To appear in ``Magnetic Coupling between the Interior 
       and the Atmosphere of the Sun'', eds. S.~S.~Hasan and R.~J.~Rutten, 
       Astrophysics and Space Science Proceedings, Springer-Verlag, 
       Heidelberg, Berlin, 2009.} \vspace*{-5ex} \par
 \def\lastand{\ifnum\value{@inst}=2\relax
                 \unskip{} \andname\
              \else
                 \unskip \lastandname\
              \fi}%
 \def\and{\stepcounter{@auth}\relax
          \ifnum\value{@auth}=\value{@inst}%
             \lastand
          \else
             \unskip,
          \fi}%
  \raggedright
 {\Large \bfseries\boldmath
  \pretolerance=10000
  \let\\=\newline
  \raggedright
  \hyphenpenalty \@M
  \interlinepenalty \@M
  \if@numart
     \chap@hangfrom{}
  \else
     \chap@hangfrom{\thechapter\thechapterend\hskip\betweenumberspace}
  \fi
  \ignorespaces
  \@title \par}\vskip .8cm
\if!\@subtitle!\else {\large \bfseries\boldmath
  \vskip -.65cm
  \pretolerance=10000
  \@subtitle \par}\vskip .8cm\fi
 \setbox0=\vbox{\setcounter{@auth}{1}\def\and{\stepcounter{@auth}}%
 \def\thanks##1{}\@author}%
 \global\value{@inst}=\value{@auth}%
 \global\value{auco}=\value{@auth}%
 \setcounter{@auth}{1}%
{\lineskip .5em
\noindent\ignorespaces
\@author\vskip.35cm}
 {\small\institutename\par}
 \ifdim\pagetotal>157\p@
     \vskip 11\p@
 \else
     \@tempdima=168\p@\advance\@tempdima by-\pagetotal
     \vskip\@tempdima
 \fi
}
\makeatother     
\fi


\title*{The Evershed Effect with SOT/Hinode}


\author{K. Ichimoto\inst{1}
        and
        the SOT/Hinode Team}

\authorindex{Ichimoto, K.} 


\institute{Kwasan and Hida Observatories, Kyoto University, Japan}

\maketitle

\setcounter{footnote}{0}  

\begin{abstract} 
The Solar Optical Telescope onboard Hinode revealed the fine-scale
structure of the Evershed flow and its relation to the filamentary
structures of the sunspot penumbra. The Evershed flow is confined in
narrow channels with nearly horizontal magnetic fields, embedded in
a deep layer of the penumbral atmosphere.  It is a dynamic phenomenon with
flow velocity close to the photospheric sound speed.
Individual flow channels are associated with tiny upflows of hot gas
(sources) at the inner end and downflows (sinks) at the outer end.
SOT/Hinode also discovered ``twisting'' motions of penumbral
filaments, which may be attributed to the convective nature of the
Evershed flow.  The Evershed effect may be understood as a natural
consequence of thermal convection under a strong, inclined magnetic
field.  Current penumbral models are discussed in the lights of
these new Hinode observations.
\end{abstract}

\section{Introduction}      \label{ichimoto-sec:introduction}

Since its discovery by \citet{ichimoto-Evershed_1909}, the
Evershed effect has been one of the longstanding mysteries in solar
physics.  It is evident from recent high-resolution observations that
the flow is closely related to the small-scale filamentary structures
in penumbrae (for reviews see \cite{ichimoto-Solanki_2003},
\cite{ichimoto-Thomas_2004}, 2008,  \cite{ichimoto-BellotRubio_2007},
\cite{ichimoto-Thomas_2009}).
The inclination of the penumbral magnetic field fluctuates 
in the azimuthal direction with the spatial scale of the penumbral filaments
(interlocking comb structure), 
whereas the flow takes place in radial filaments which have nearly
horizontal magnetic field
(\cite{ichimoto-Degenhardt_1991}, \cite{ichimoto-Schmidt_1992}, 
\cite{ichimoto-Title_1993}).
The flow vector is parallel to the magnetic fields 
(\cite{ichimoto-BellotRubio_2004}, \cite{ichimoto-Borrero_2005}).

To account for the filamentary structure of penumbrae, 
the following models have been proposed, \ie\
the embedded or rising flux tube model 
(\cite{ichimoto-Solanki_1993}, \cite{ichimoto-Schlichenmaier_1998})
in which the Evershed flow channels are explained as rising flux tubes 
embedded in more vertical background magnetic fields in the penumbra,
the gappy penumbral model (\cite{ichimoto-Spruit_2006}, 
\cite{ichimoto-Scharmer_2006})
in which the bright penumbral filaments are regarded as manifestations
of the protrusion of non-magnetized, convecting hot gas into the
oblique background fields of the penumbra, and the downward pumping
model (\cite{ichimoto-Thomas_2002},
\cite{ichimoto-Weiss_2004}, \cite{ichimoto-Brummell_2008})
in which the penumbral fine structure is created by localized
submergence of penumbral fields that are forced by the photospheric
convection around the outer edge of penumbrae.  There is still no
consensus on the physical nature and origin of the penumbral fine
structures.

Using highly stable time sequences of sunspot images and high precision 
spectropolarimetric data provided by 
the Solar Optical Telescope (SOT, \cite{ichimoto-Tsuneta_2007}, 
\cite{ichimoto-Suematsu_2007}) onboard Hinode (\cite{ichimoto-Kosugi_2007}),
\citet{ichimoto-Ichimoto_2007a} confirmed the interlocking-comb 
penumbral structure
and found that the Evershed flow preferentially takes place
in bright filaments in the inner penumbra and in dark filaments in the
outer penumbra.  They also found the presence of a number of small
patches with a vertical velocity component, upward motions distributed
over the penumbra, and strong downward motions associated with
magnetic polarity opposite to that of the sunspot in the mid and outer
penumbra.  The vertical motions may be regarded as the sources and
sinks of the Evershed flow channels, though unequivocal identification of
individual pairs was not reached.

Here we report results on the spatial distribution of vertical motions
in penumbrae obtained from further analyses of the sunspot data taken
by SOT.  The characteristics of the Evershed flow are summarized, and
the origin of the flow is discussed within the context of the present
penumbral models.

\section{Elementary structure of the Evershed flow}   
\label{ichimoto-sec:result}

\begin{figure}  
  \centering
  \includegraphics[width=\textwidth]{\figspath/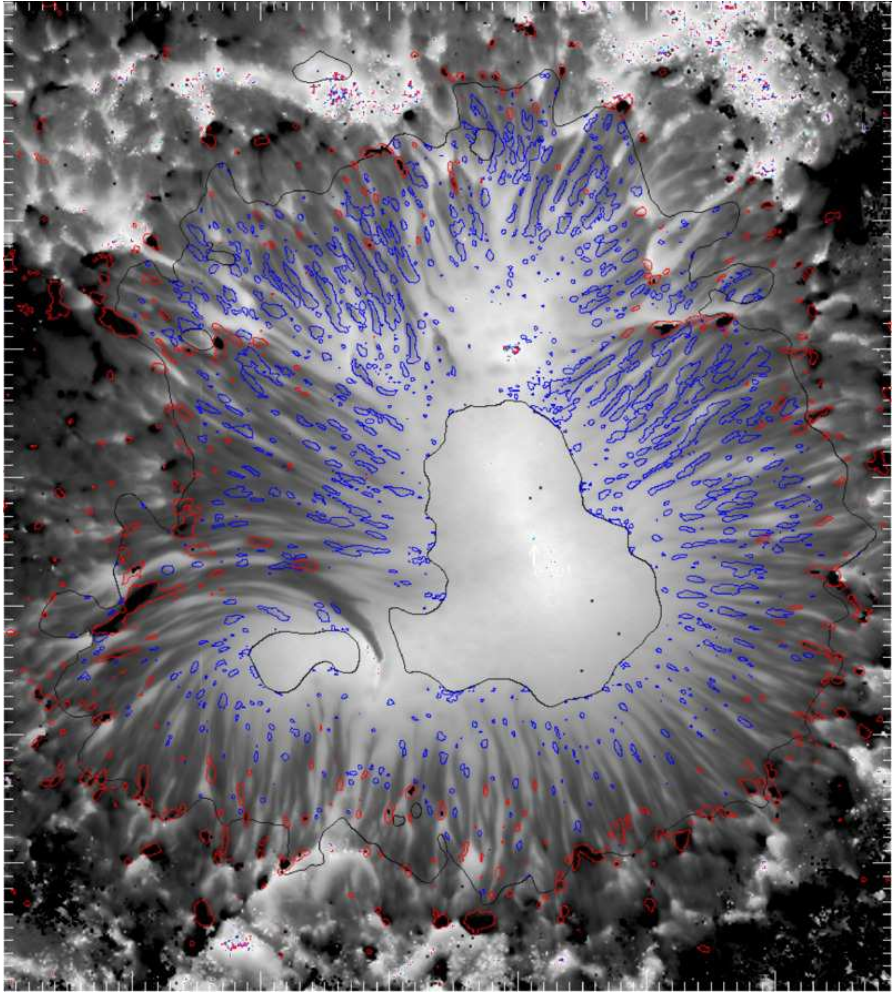}
  \caption[]{\label{ichimoto-fig:001}
%
Inclination of the magnetic field.
This sunspot was observed by SP/SOT on May 1, 2007.
Top is toward the disk center; the heliocentric distance is 5.8\deg.
The black contours show the inner and outer boundaries of the penumbra.
Dark channels in the penumbra represent nearly horizontal fields.
Overlaid are contours showing the vertical motions, i.e., blue for upflow
and red for downflow.
}\end{figure}

\begin{figure}  
  \centering
  \includegraphics[width=\textwidth]{\figspath/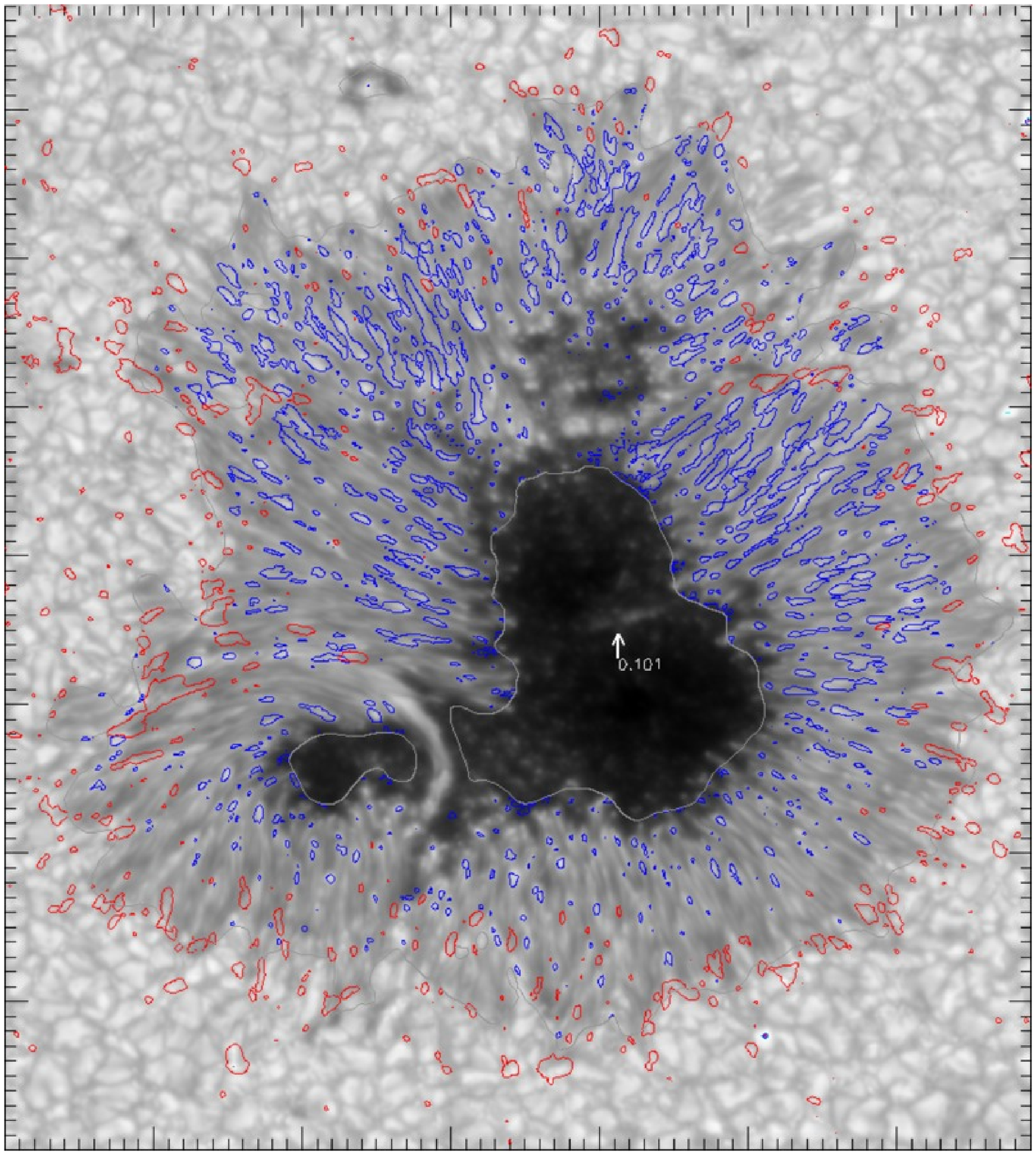}
  \caption[]{\label{ichimoto-fig:002}
%
The same sunspot in continuum intensity near $\lambda=630$~nm,
with the 
same contours as in Fig.~\ref{ichimoto-fig:001}.
}\end{figure}

The spatial distribution of the vertical motions in sunspots were
studied using SOT/SP data taken on May 1, 2007.  The sunspot was
located near disk center, with the viewing angle between the line of
sight and the normal to the solar surface $\mu=5.8$\deg.  Stokes profiles of
the pair of \FeI\ lines at 6302~\AA\ were fitted using the
Milne-Eddington model to obtain vector magnetic fields.

Figure~\ref{ichimoto-fig:001} shows the magnetic field inclination
overlayed with contours showing the vertical motions.  The blue contours
show blueshift at $-0.8$\,km\,s$^{-1}$ in the wing of Stokes-$I$ of
\FeI\ 6301.5\,\AA\ and the red countours show $V/I_c = 0.01$ at \FeI\
6302.5\,\AA\ + 0.365\,\AA, representing strong downflow regions with
opposite magnetic polarity to the spot.  $V$ and $I_c$ are Stokes-$V$
and the continuum intensity, respectively.  Radial penumbral filaments
with dark appearance in the inclination map are the channels with
nearly horizontal magnetic field.  It is remarkable that the upflow
and downflow patches are aligned with the horizontal field channels
that carry the Evershed flow, and that small-scale upflows are
preferentially located near the inner ends, downflows at the outer
ends of the horizontal field channels.  Thus, the upflow and downflow
patches may be regarded as the sources and sinks of the Evershed flow
elements, embedded in the deep penumbral photosphere.  It is noted
that the Stokes-V signal is well visible even at \FeI\ 6302.5\,\AA\ +
0.454\,\AA\ (22\,km\,s$^{-1}$). This suggests the presence of
supersonic flows near the outer end of the flow channels (see
\cite{ichimoto-Ichimoto_2007a}).

Figure~\ref{ichimoto-fig:002} shows the continuum intensity of the
sunspot overlayed with the same contours as in
Fig.~\ref{ichimoto-fig:001}.  The spatial correlation between bright
grains and upflows is remarkable.  It is obvious that the Evershed
(up-)flow carries the energy to maintain the penumbral brightness,
although quantitative evaluation of the energy flux carried by the
upflow is not straightforward because the line forming layer is likely
well above the vigorously convectiving layer.

\section{Discussion}                   \label{ichimoto-sec:discussion}

\citet{ichimoto-Ichimoto_2007b} found a ``twisting'' motion
in the leading part of the bright filaments in penumbrae.
Since the visibility of this twisting motion depends  on 
the viewing angle, i.e., it is seen only when the penumbral
filaments are observed from their side, 
\citet{ichimoto-Ichimoto_2007b} interpreted
the apparent twisting feature as a manifestation of overturning convection
in the source region of the Evershed flow.
\citet{ichimoto-Zakharov_2008} confirmed the twisting motions of 
penumbral filaments 
with ground-based observations and concluded that the twisting
features indicate the presence of convective rolls in filaments with a
nearly horizontal magnetic field.

From the analysis of the net circular polarization in penumbrae,
\citet{ichimoto-Ichimoto_2008} found evidence of positive correlation 
between the Evershed flow velocity and the magnetic field strength,
spatially along the line of sight, in the sense that the flow velocity
and the magnetic field strength both increase with depth in the flow
channels.  This is supports the conclusion that the flowing gas is
strongly magnetized.  The Stokes-$V$ Dopplershift or the Dopplershift
of polarized spectral-line components in penumbrae serves as more
direct evidence that the flowing gas is not field-free.

The properties of the Evershed effect so far obtained are summarized
as follows:

\begin{itemize} \itemsep=1ex 

  \item Each Evershed flow channel is associated with upflow (source) 
	and downflow (sink) in its inner and outer ends;

  \item The geometry of the magnetic field is consistent with low-lying 
	arched tubes embedded in more inclined magnetic fields,
	i.e., so-called interlocking-comb penumbral structure;

  \item There is a very good spatial correlation between upflows
  and bright penumbral grains, which strongly suggests that the
  Evershed (up-)flow carries the energy for maintaining the penumbral
  brightness;

  \item The source region of the Evershed flow channels shows hints of 
	overturning convection;

  \item The flow velocity (and the magnetic field strength) increase 
	with depth in the visible layers of the flow channel;

  \item The flow plasma is not field-free but magnetized.

\end{itemize}

These characteristics obviously suggest a convective origin for the
Evershed effect.  Let us consider the currently representative
penumbral models, i.e., the embedded-fluxtube model and the gappy
model, in the light of these observational results.  There is no
observational evidence of the presence of a lower boundary of
fluxtubes at least in the inner and middle penumbra.  Thus, the
picture of narrow fluxtubes in the embedded-fluxtube model is not
supported by observation.  On the other hand, the flowing gas is
highly magnetized, which is an obvious contradiction between the
field-free gap model and the observations.  If the fluxtube model
allows vertically elongated ``flux tubes'' (or slabs), and if the gap
model discards the term ``field free'', then there is no fundamental
difference between the two models.  In both models, rising motions of
hot gas are driven by buoyancy forces.  These new observational
results suggest that the Evershed effect can be understood as a
natural consequence of thermal convection under strong, inclined
magnetic fields.

Recent 3D numerical MHD simulations (\cite{ichimoto-Heinemann_2007}, 
\cite{ichimoto-Rempel_2008}) 
of sunspots have successfully reproduced
the interlocking-comb structure of the magnetic fields in penumbrae
with Evershed outflows along filaments with nearly horizontal 
magnetic field and overturning motions.
\citet{ichimoto-Scharmer_2008} argued that the Evershed flow represents 
the horizontal flow component of overturning convection in penumbrae,
based on numerical simulations.  The velocity of the Evershed outflow
in current numerical simulations is, however, only a few km\,s$^{-1}$,
which is much smaller than the observed one.  The mechanism that
drives the vigorous observed flow, especially at the outer end of the
flow channel, is still to be identified.  Siphon flow models (e.g.,
\cite{ichimoto-Thomas_1992}) predict supersonic flows near the outer
footpoint of the flow channel (\cite{ichimoto-Montesinos_1997}),
although the theory is based on a stationary model.  A more dynamic
picture is given by the downward pumping model
(\cite{ichimoto-Weiss_2004}), in which stochastic flows may be driven
by the convective collapse caused by the localized submergence of
penumbral magnetic fields forced by the photospheric convection around
the outer edge of penumbrae.  The validity of these various models
should be examined by investigating the temporal evolutions of the
individual Evershed flow elements in further observations.

\begin{acknowledgement}
  Hinode is a Japanese mission developed and launched by ISAS/JAXA,
  with NAOJ as domestic partner and NASA (USA) and STFC (UK) as
  international partners. It is operated by these agencies in
  co-operation with ESA and NSC (Norway).  This work was carried out
  at the NAOJ Hinode Science Center, which is supported by the
  Grant-in-Aid for Creative Scientic Research, The Basic Study of
  SpaceWeather Prediction from MEXT, Japan (Head Investigator:
  K. Shibata), generous donations from Sun Microsystems, and NAOJ
  internal funding.
\end{acknowledgement}

\begin{small}



\end{small}

\end{document}